\documentclass{article}

\usepackage{microtype}
\usepackage{graphicx}
\usepackage{subfigure}
\usepackage{booktabs} 
\usepackage{multirow}

\usepackage{hyperref}

\usepackage[accepted]{icml2020}
\usepackage{amsmath,amsfonts,amssymb,dsfont}

\def\G{\mathcal{G}}

\def\D{\mathcal{D}}

\def\our{LocoGAN}

\icmltitlerunning{LocoGAN -- Locally Convolutional GAN}

\begin{document}

\twocolumn[
\icmltitle{ LocoGAN -- Locally Convolutional GAN }



\icmlsetsymbol{equal}{*}

\begin{icmlauthorlist}
\icmlauthor{\L{}ukasz Struski}{uj}
\icmlauthor{Szymon Knop}{uj}
\icmlauthor{Jacek Tabor}{uj}
\icmlauthor{Wiktor Daniec}{uj}
\icmlauthor{Przemys\l{}aw Spurek}{uj}
\end{icmlauthorlist}

\icmlaffiliation{uj}{Faculty of Mathematics and Computer Science, Jagiellonian University, Krak\'ow, Poland}

\icmlcorrespondingauthor{Przemys\l{}aw Spurek}{przemyslaw.spurek@uj.edu.pl}


\icmlkeywords{Machine Learning, ICML}

\vskip 0.3in
]



\printAffiliationsAndNotice{\icmlEqualContribution} 

\begin{abstract}

In the paper we construct a fully convolutional GAN model: LocoGAN, which latent space is given by noise-like images of possibly different resolutions. The learning is local, i.e. we process not the whole noise-like image, but the sub-images of a fixed size.
As a consequence LocoGAN can produce images of arbitrary dimensions e.g. LSUN bedroom data set. 
Another advantage of our approach comes from the fact that we use the position channels, which allows the generation of fully periodic (e.g. cylindrical panoramic images) or almost periodic ,,infinitely long'' images (e.g. wall-papers).

\end{abstract}


\section{Introduction}

Generative adversarial networks (GANs) \cite{goodfellow2014generative} are one of the most important areas of deep learning. 
GANs based on deep convolutional networks \cite{radford2015unsupervised,karras2017progressive,zhang2018stackgan++}  have been especially successful. Standard GAN models generate new images from a noise of fixed dimension, and consequently produce images of fixed resolution. 

We present a new architecture and training method for GANs that are aware of spatial information\footnote{ The code is available \url{https://github.com/gmum/LocoGAN} }. An important feature of our model is its simplicity and applicability to most standard GAN models.

The key idea consists of four elements:
\begin{itemize}
\setlength\topsep{0pt}
\setlength\partopsep{0pt}
\setlength{\parskip}{0pt}
\setlength{\itemsep}{3pt plus 1pt}
\item We use a fully convolutional architecture for generator network.
\item Latent space consists of noise images with potentially arbitrary resolution and a fixed number of channels.
\item We divide the latent noise into local and global noise.
\item We add extra channels with spatial information to the input noise images.
\end{itemize}
Such architecture and design of latent space allows us to use an input of various dimensions. We use that to train our model only on parts of the latent image, see Fig.~\ref{fig:shemma}. We call this approach  {\em local learning}. Section 3 contains the detailed description of the model and the training procedure.


\begin{figure}[htb]
\begin{center}
\includegraphics[width=0.48\textwidth]{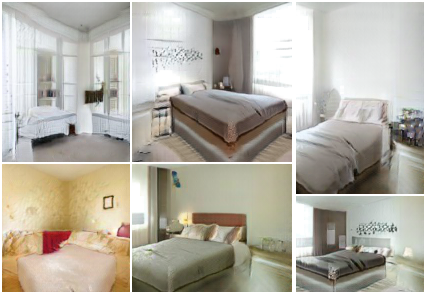}\\
\end{center}
	\caption{The figure presents samples of different resolutions generated by a model trained on LSUN (bedrooms) data set. Resolutions of images in the first row from left to right are 
$160\times128$, $160\times160$, $192\times128$ and $128\times128$, $128\times160$, $96\times128$, respectively for the second row. }
	\label{fig:lsun_resize}
\end{figure}
Despite the fact entire images are never used in training, \our{} produces full,  state-of-the-art-quality images. We further demonstrate (see the following examples and Section~\ref{sec:exp}) a variety of novel applications enabled by teaching the network to be aware of coordinates as well as local/global latent. In Fig.~\ref{fig:lsun_resize} we present samples with different sizes, generated by a model trained on LSUN (bedrooms) data set.

The next important consequence of local learning is that we can train \our{} just on one source image -- under the assumption that the image has a strong pattern-like structure. As a consequence by choosing periodic position channels, \our{} produces periodic pattern images that match the structure of the source image.

Another interesting consequence of our model is that since our model is aware of the position of a given feature, we can exchange some features between images just by exchanging the corresponding parts in the respective noise-like images. For example we can trivially ,,transplant'' a part of one image (like smile or glasses) to the other one, see Fig.~\ref{fig:celeba_attr}.

Let us now briefly describe the content of the paper. In the next section we provide the related works. In the third section we provide the description of our model. Fourth section provides the basic experiments on our model (for more detailed experiments see the Appendix).
We conclude the paper with the short conclusion. 

\begin{figure}[htb]
\begin{center}
    \includegraphics[width=0.45\textwidth]{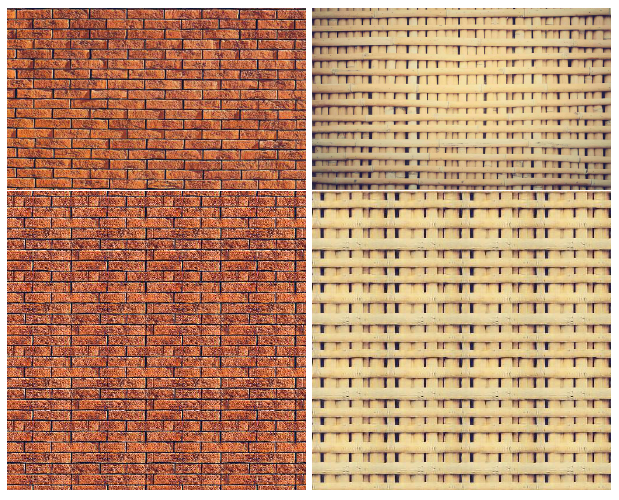}
\end{center}    
    \caption{In the experiment we take one picture of a high resolution. Based on that image we create a data set, by cropping squares of fixed resolution ($192\times192$). We train our model to generate periodic parts. We present original images (the first row) and image generated by \our{}. The image consists of one fragment repeated many times.}
\label{fig:peridic_int}
\end{figure}

\begin{figure}[htb]
	\begin{center}
	\includegraphics[width=0.45\textwidth]{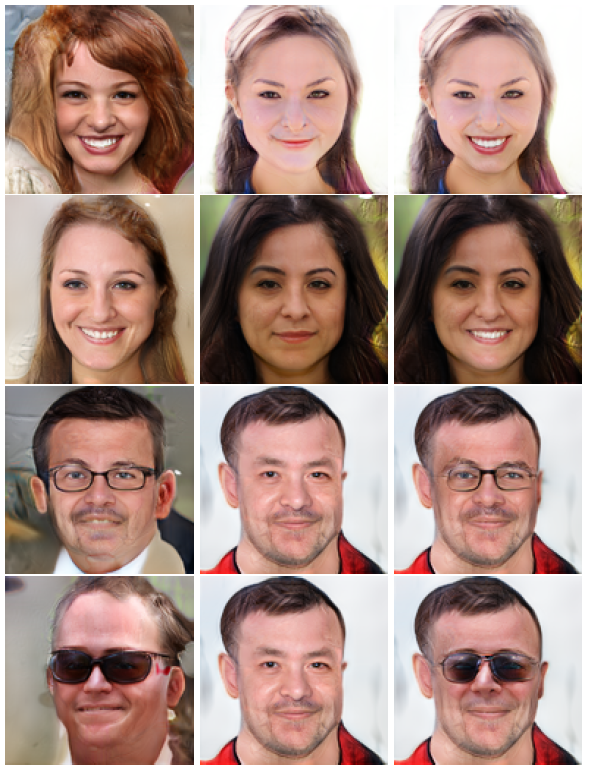}
	\end{center}
	\caption{Moving spatially localized attributes from one image to another. In the first two rows we add smile from the first image to the second one. In the next two rows we add glasses from the first image to the second one. Due to the fact that our model is local, we do not change the image except for a neighbourhood of the ,,transplanted'' positions.}
	\label{fig:celeba_attr}
\end{figure}




\section{Related works}

Generative modeling is a broad area of machine learning which deals with modeling a joint distribution of data. Roughly speaking, we want to produce examples similar to the ones already in a data set $X$, but not the same.

Generative models are one of the fastest growing areas of deep learning. In recent years a number of generative models, like Variational AutoEncoders (VAE) \cite{kingma2014auto}, Wasserstein  AutoEncoder  (WAE) \cite{tolstikhin2017wasserstein},  generative adversarial networks (GAN) \cite{goodfellow2014generative}, auto-regressive models \cite{isola2017image} and flow-based generative models \cite{dinh2014nice,kingma2018glow}, were constructed.

The quality of generative image modeling has increased in recent years thanks to  Generative Adversarial Networks (GANs) architecture.
GANs have solved many problems in various image generation tasks like
image-to-image translation \cite{isola2017image,zhu2017unpaired,taigman2016unsupervised,park2019semantic}, image super-resolution \cite{ledig2017photo,sonderby2016amortised} and text-to-image synthesis \cite{reed2016generative,hong2018inferring}.

GAN is a framework for training deep generative models using a mini-max game. The goal is to learn a generator distribution $P_{\G}(x)$ that matches the real data distribution $P_{data}(x)$.  GAN learns a generator network $\G$ that generates samples from the generator distribution $P_{\G}$ by transforming a noise variable $z \sim P_{noise}(z)$ (usually Gaussian noise $N(0,I)$) into a sample $\G(z)$. This generator is trained by playing against an adversarial discriminator network $\D$ that aims to distinguish between samples from the true data distribution $P_{data}$ and the generator’s distribution $P_{\G}$. More formally, the minimax game is given by the following expression:

$$
\begin{array}{c}
\min_{\G} \max_{\D} V(\D,\G) = \\[1em]
\mathbb{E}_{x \sim P_{data}} [\log \D(x)] +  \mathbb{E}_{x \sim noise} [\log (1-\D(\G(x)))].
\end{array}
$$

The main advantage over other models is the ability to produce sharp images, which are indistinguishable from real ones. 
GANs are impressive in terms of the visual quality of images sampled from the model, but the training process is often hard and unstable.

In recent years many researcher focused on modifications to the vanilla GAN procedure to improve stability of the training process.

{\bf Changing the objective function}
The first part of such modification is based on changing the
objective function \cite{arjovsky2017wasserstein,mao2017least,bellemare2017cramer} to encourage convergence. 

In \cite{arjovsky2017wasserstein} authors introduced Wasserstein Generative Adversarial Networks (WGAN) which is an
alternative to vanilla GAN. 
Instead of using a discriminator to distinguish real and fake images (generated by GAN model), the WGAN replaces the discriminator model with a critic that scores the realness of a given image.
WGAN improves the stability of learning and partially solves problems with mode collapse.

{\bf Restrictions on the gradient penalties}
Another approach use some restrictions on the decoder gradient penalties \cite{gulrajani2017improved,kodali2017convergence}. 
The most promising approach in this area is Spectral Normalization \cite{miyato2018spectral}, which uses spectral norm of the weight matrices in the discriminator in order to constrain the Lipschitz constant of the discriminator function.
In \cite{zhang2018self} authors use spectral normalization also for the generator. Spectral normalization in
the generator can prevent the escalation of parameter magnitudes and avoid unusual gradients. 
In practice it allows to  stabilize the GAN training dynamics \cite{miyato2018cgans}.

{\bf Imbalanced learning rate for generator and discriminator}
In literature it is well known that regularization of the discriminator \cite{gulrajani2017improved,miyato2018spectral} often slows down the GANs’ learning process. In practice, when we use regularized discriminators we require more updates for the discriminator than for the generator. 
We can solve this problem by using individual learning rate for both the
discriminator and the generator \cite{heusel2017gans}.
Such solution speeds up the convergence of GAN models drastically.

{\bf Self-Attention mechanisms } 
Recently, attention mechanisms have
been successfully applied in many different areas of deep learning \cite{bahdanau2014neural,parmar2018image,vaswani2017attention,xu2015show}.
In particular, self-attention \cite{cheng2016long,parikh2016decomposable}, also called intra-attention, calculates the response at a position in a sequence by attending to all positions within the same sequence. 

SAGAN (Self-Attention Generative Adversarial Network)~\cite{zhang2018self} learns to efficiently find global, long-range dependencies within internal representations of images. In addition, the self attention mechanism in GAN can be combined with spectral normalization to stabilize GAN traning. The imbalanced learning rate for the generator and the discriminator speeds up the training.

{\bf  Progressive growing architecture } 
The recent extension ProGAN \cite{karras2017progressive} trains high-resolution GANs in the single-class setting by training a single model across a sequence of increasing resolutions. ProGAN progressively extends the generator and the discriminator architecture, starting from easier low-resolution images, and adding new layers that introduce higher-resolution details as the training progresses. Style-Based Gan (styleGan) \cite{karras2019style} is a modification of ProGAN with use of alternative generator architecture, borrowing from style transfer literature.

{\bf Coordinates of pixels in GAN model }
The recent extension of GANs architecture improves the stability of a model but still there is a problem with training GANs on images with varying resolution.
In \cite{lin2019coco} authors present Conditional Cordinate GAN
(COCO-GAN) new architecture where generator generates images by
parts based on their spatial coordinates as the condition. On
the other hand, the discriminator learns to justify realism
across multiple assembled patches by global coherence, local appearance, and edge-crossing continuity. Although full images are never generated during training, COCO-GAN can produce full images.  Furthermore, COCO-GAN can produce images that are larger than training samples (“beyond-boundary generation”).
The use of coordinates allows to generate panoramic images by using cylindrical coordinate system.

COCO-GAN is very close to our model in terms of possible applicability.  Both of models use coordinate of pixels and can produce periodic images. On the other hand, contrary to COCO-GAN our model use fully convolutional architecture and global coordinates. The architectures are completely different but can be used in similar non trivial task. For instead, our model can be train with images with different resolution and COCO-GAN can be used to extrapolation of images.   



\section{\our{} main idea}

\begin{figure}[htb]
\begin{center}
\includegraphics[height=4.0cm]{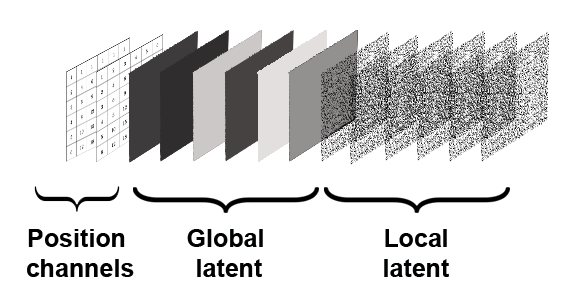}\\
\end{center}
  \caption{Construction of latent noise-like image. Global latent represents global features and is the same for every pixel for a given channel in an image. Local latent is a gaussian noise without any correlation between pixels.}
  \label{fig:latent}
\end{figure}

Our approach uses the fact that a fully convolutional neural network (generator) can generate images of any resolution. The larger input we use, the larger output we get. In \our{} we train only on small parts of images, where the input consists of image-like noise of arbitrary size, see Fig.~\ref{fig:latent}. After training we can generate images of different resolutions by increasing the size of input noise layer. 
This means that \our{} can be trained on a data set containing different resolutions like LSUN bedroom scene.  

Since our approach is a simple modification of the standard approach, let us briefly describe it here.

{\bf Fully convolutional architecture.}
Convolutional neural networks are built on translation invariance.
Each layer of data in a convolutional layer is a three-dimensional array of size $h \times w \times d$, where $h$ and $w$ are spatial dimensions, and $d$ is the feature or channel dimension. The first layer is the image, with pixel size $h \times w$, and $d$ color channels. In the case of generator, as an input we use an array of size $h \times w \times d$ containing a Gaussian blur. Fully convolutional architecture can process images of arbitrary resolution (only the number of color channels is fixed).

\begin{figure}[htb]
\begin{center}
\includegraphics[height=8.5cm]{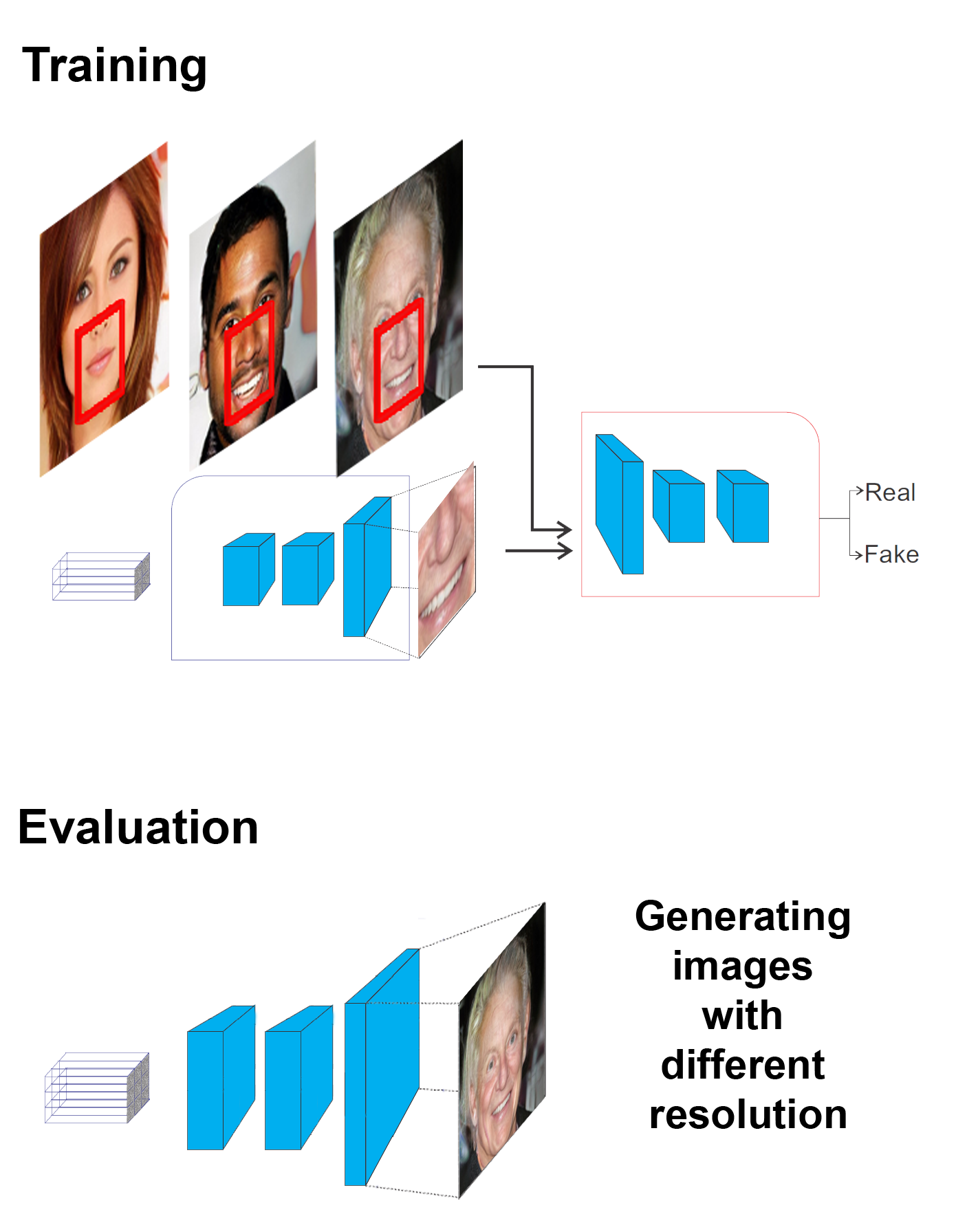}\\
\end{center}
  \caption{In this figure we explain the idea of local learning. We train model on image with possible different resolutions by using small cropped part of size ($64\times64$). We append additional coordinate channels to both generator's and discriminator's input to inform the model which part of the image it is working on. This allows us to generate images of different resolutions by manipulating the generator's input image size.}
  \label{fig:shemma}
\end{figure}



To construct \our{}, locally convolutional GAN, one has to deal with a few problems. First, since the discriminator has a fixed dimension, we apply the idea of {\em local learning}. That is during training we always process the subimages of fixed size.
\begin{figure*}[htb]
	\centering
	\includegraphics[width=0.95\textwidth]{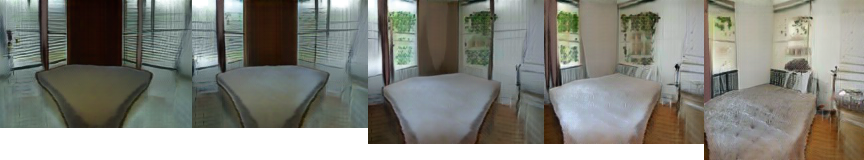}
	\caption{Linear interpolation between images with different resolutions:  endpoints are the size from $128\times192$ to $160\times160$, respectively.}
	\label{fig:lsun_interpolation}
\end{figure*}

\begin{figure*}[htb]
	\centering
 	\includegraphics[width=0.8\textwidth]{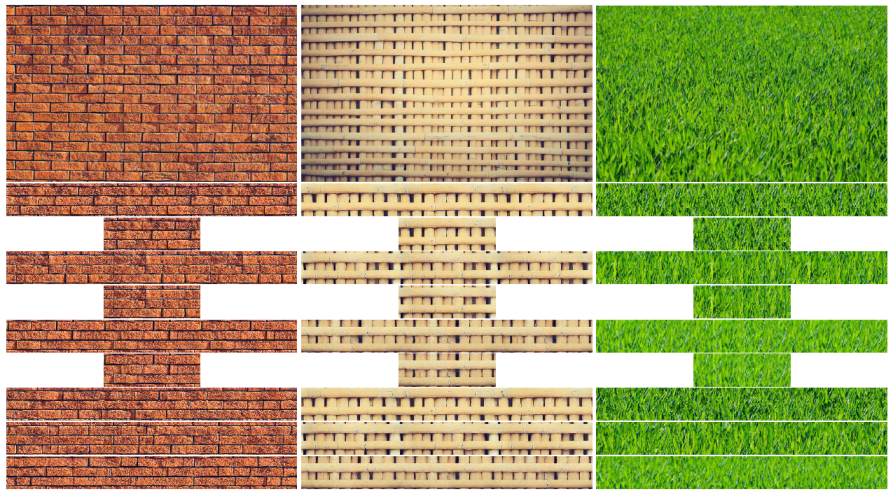}

\caption{In the experiment we take one picture of a high resolution. Based on that image we create a data set, by cropping rectangles of fixed resolution ($64\times192$). We train our model to generate periodic parts. We present original images (the first row), elements generated by our model (we highlight the part which is periodic) in the next six rows. In the last three rows we show that we can change local plan to achieve a small modification of each periodic element.}
\label{fig:peridic_2}
\end{figure*}

\paragraph{Adding spatial information: position channels.}
Since we have fully convolutional architecture, the model (for example for the FFHQ like data, where the global structure is crucial), needs the information that it should construct the neighbourhood of the nose, or of the eye. In the classical GAN models, the model derives this information from the fact that the convolutions are used  with zero padding, and it can easily deduce the relative position from the boundary. Since in our model we do not allow the zero padding, the images do not have boundary which can be recognized by the network. Instead, we include the information about the position in additional channels. In the simplest basic model we add two channels with $x$ and $y$ positions of the pixel, see Fig.~\ref{fig:latent}, however, to construct periodic images we use more channels (a basic form for $x$ periodic images is given by $[\cos x, \sin x]$). In our model we add the position channels at every layer of the network. Since each layer has different size $h \times w $ we have to resize channels with coordinates respectively to the size $h \times w$. We apply classical downsampling  of original channels with coordinates to smaller resolution.  

{\bf Padding.} As a consequence of the above described approach our model can never see the end of an image (understood as pixels which originally did not belong to the input noise-like image). Therefore during the training process we crop larger part of an image and use additional noise pixels as a padding (a simplified version of this approach is simply no padding). The size of the noise padding must be large enough to produce fragment of an image with size equal to target cropped element.  

\paragraph{Global and local latent.}
However, we still need one additional element. Since our construction is fully convolutional, we have no guarantee that, for example, the left upper local corner of the input would be able to exchange information with the right lower. In consequence, for the FFHQ data set we could theoretically obtain a photo where the left side of the image comes from a man, while the right side comes from a woman. Thus, in our model we need to model local representation of an image as well as the entire image plan. The final generator must have the ability to produce images with higher resolution than elements in the training set. Therefore, we have to divide latent (input noise layer in generator) into two parts. The first one for coding global feature of the training data set which is fixed during training for all elements. And the second one for coding local representation of the image which is different for each element from the data set.      
The global (master plan) latent encodes global information from image and the local one is responsible for the local structure. 

Concluding, in the global latent channels we input image-type noise which is constant over each channel, see Fig.~\ref{fig:latent}. To allow the model some additional flexibility, we also add {\em local latent} where we sample by  gaussian noise. In practice the local plan corresponds to some local features (like a strand of hair changing its position),~see~Fig.~\ref{fig:celeba-random_local}.

\paragraph{Architecture.}
Our architecture derived from the DCGAN model \cite{radford2015unsupervised}. To construct a generator we use the convolutional-transpose, and batch normalization with ReLU activation function for all layers except for the last one. For the final layer we use Tanh without batch normalization. Detailed values of parameters such as input channels (in\_ch), output channels (out\_ch), kernel size, stride and padding in individual layers are presented in Tab~\ref{tab:arch_1}.

\begin{table}[]
\begin{center}
\begin{tabular}{c@{\hskip 6pt}r@{\hskip 6pt}r@{\hskip 6pt}r@{\hskip 6pt}r@{\hskip 6pt}r@{\hskip 0pt}}
No. layer & in\_ch & out\_ch & kernel & stride & padding \\ \toprule
1 & 3 & 1024 & 4 & 2 & 3 \\ 
2 & 1024 & 512 & 4 & 2 & 3 \\
3 & 512 & 256 & 4 & 2 & 3 \\
4 & 256 & 128 & 4 & 2 & 3 \\
5 & 128 & 3 & 4 & 1 & 3
\end{tabular}
\end{center}
\caption{Detailed values of parameters used in generator.}
\label{tab:arch_1}
\end{table}

As input to the discriminator we take a $3\times64\times64$ input image, processes it through five convolution and LeakyReLU layers, and output the final probability through a Sigmoid activation function. For a stable training model in the first four layers we used spectral normalization \cite{miyato2018spectral} before activation function.  Detailed values of parameters such as input channels (in\_ch), output channels (out\_ch), kernel size, stride and padding in individual layers are presented in Tab.~\ref{tab:arch_2}.

\begin{table}[]
\begin{center}
\begin{tabular}{c@{\hskip 6pt}r@{\hskip 6pt}r@{\hskip 6pt}r@{\hskip 6pt}r@{\hskip 6pt}r@{\hskip 0pt}}
No. layer & in\_ch & out\_ch & kernel & stride & padding \\ \toprule
1 & 3 & 64 & 4 & 2 & 1 \\ 
2 & 64 & 128 & 4 & 2 & 1 \\
3 & 128 & 256 & 4 & 2 & 1 \\
4 & 256 & 512 & 4 & 2 & 1 \\
5 & 512 & 1 & 4 & 1 & 0
\end{tabular}
\end{center}
\caption{Detailed values of parameters used in discriminator.}
\label{tab:arch_2}
\end{table}

\section{Experiments} \label{sec:exp}

In this section we present the experiments. We start with the standard LSUN database -- we show that \our{} allows interpolation between images of different resolutions. Next we proceed to the experiments where we show that we can learn from only one pattern image. In the last subsection we study the role of channels on the faces data set.

In all experiments on LSUN and FFHQ we train our architecture on small cropped part ($64 \times 64$) of original image ($128 \times 128$). 
During training of the model we use image-type noise input with 16 channels for master plan, 2 channels for local plane, and 2 position channels with size $10\times10$. Positions of original image ($128 \times 128$) are scaled to [-1,1].

\subsection{LSUN}

The LSUN model was trained with the default parameters. The position channels were given by
renormalization of the position of a given pixel, i.e.:
\begin{itemize}
    \setlength\topsep{0pt}
    \setlength\partopsep{0pt}
    \setlength{\parskip}{0pt}
    \setlength{\itemsep}{0pt plus 5pt}
    \item X-position: $x$,
    \item Y-position: $y$.
\end{itemize}

In the case of LSUN \cite{yu2015lsun} (bedroom) we scale shorter edge of the image to $128$ and train model on images with different resolutions.  
Our model is trained on  ($64 \times 64$) randomly cropped elements of original data set.

\begin{figure}[htb]
	\centering
	\includegraphics[width=0.47\textwidth]{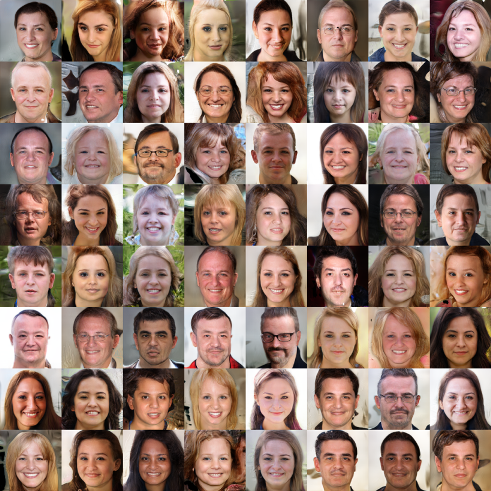}
	\caption{Samples with resolution $128\times128$ generated from model trained on FFHQ data set. FID score equals 26.3.}
	\label{fig:celeba_samples}
\end{figure}

\begin{figure*}[htb]
	\centering
	\includegraphics[width=1.\textwidth]{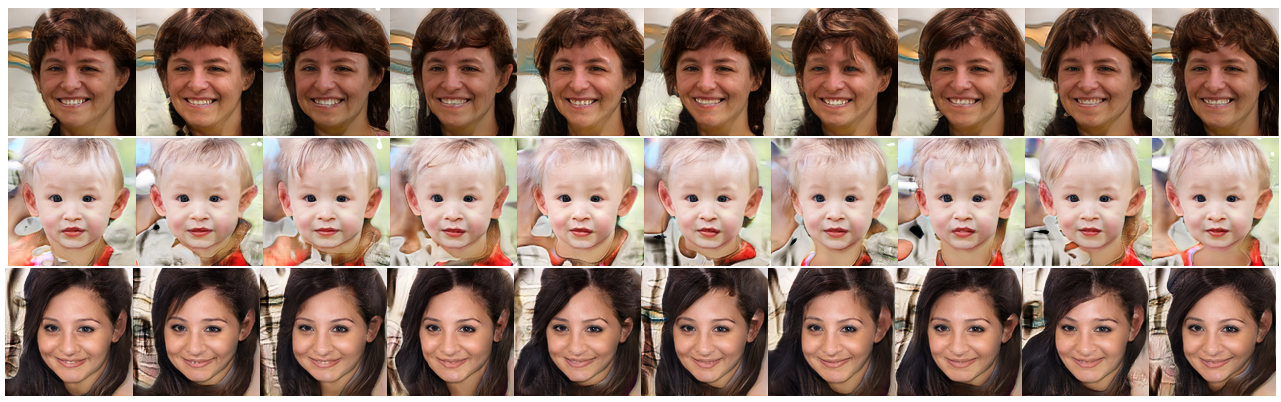}
	\caption{Samples from resolution $128\times 128$ with the same global plan, and random local plan. The picture shows that the modification of a local plan has a small effect on the image. For instance, we may notice a slight change in the hair.}
	\label{fig:celeba-random_local}
\end{figure*}

In Fig.~\ref{fig:lsun_resize} we present images  generated by \our{} model. As we can see our model produces state of the art images with the same resolution as in the training data set. 

As it was mentioned, the position channels are responsible for the position of the elements, and in particular for the up-down, left-right information (i.e. carpet is on the down of the LSUN image, while the lamp is under the ceiling).

This means, that we can reasonably interpolate between the images of different resolutions. In the case of classical interpolation we take two points from prior distribution and apply linear interpolation. Then we transfer points through the generator. In our model image-type noise layer has coordinates channels. 
Therefore, we can use sizes of target images (potentially different ones) and  simultaneously change resolution of an image (linear interpolation between sizes) and modify array of coordinates with bilinear interpolation. In Fig.~\ref{fig:lsun_interpolation} we present interpolation between two elements with different sizes.  

\subsection{Patterns}

The case of patterns is probably the most novel from features of \our{}. In this case our model learns to produce an image which is similar to the one source image.
We assume here that we have only one source image, which has a pattern-like structure. For this task we chose three images: brick $666\times1000$, bamboo $1064\times708$ and grass $821\times544$, see Fig.~\ref{fig:peridic_2}. 

In the experiment we take one picture of a high resolution. Based on that image we create a data set, by cropping rectangles of fixed resolution ($64\times192$). The cropped images should have similar distribution. Elements from training data set should be large enough to represent local structure. On the other hand, elements
should not be too large because we do not want to encode the global structure of the image.


Now, we train our model to be able to produce periodic parts. More specifically, we produce a small element that can be repeated to construct the full image, see Fig.~\ref{fig:peridic_2}.

Because the network is fully convolutional, a periodic pattern can be generated by changing position channels in our model. We use two position channels, one
for obtaining periodic strip: 
\begin{itemize}
    \setlength\topsep{0pt}
    \setlength\partopsep{0pt}
    \setlength{\parskip}{0pt}
    \setlength{\itemsep}{3pt plus 1pt}
    \item X-position: $(\cos \alpha x, \sin \alpha x)$,
    \item Y-position: $y$,
\end{itemize}
and the other one for obtaining double periodic covering of the plane:
\begin{itemize}
    \setlength\topsep{0pt}
    \setlength\partopsep{0pt}
    \setlength{\parskip}{0pt}
    \setlength{\itemsep}{3pt plus 1pt}
    \item X-position: $(\cos \alpha x, \sin \alpha x)$,
    \item Y-position: $(\cos \alpha y, \sin \alpha y)$,
\end{itemize}
see Fig.~\ref{fig:peridic_int}.
\begin{figure}[htb]
\centering
\includegraphics[width=0.23\textwidth]{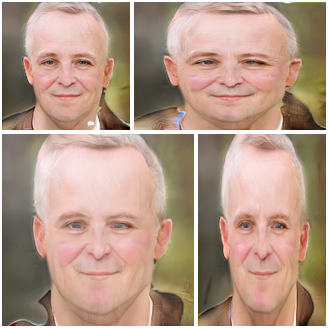}
\includegraphics[width=0.23\textwidth]{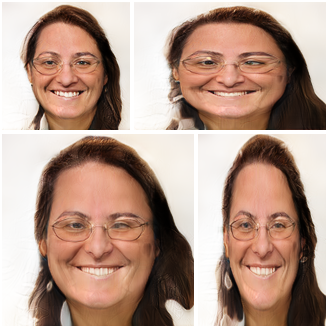}
\caption{
In the left upper corner, we present an image generated in the same resolution as the element in the training data set. \our{} can produce images with different resolutions than elements from the training data set. We present images with increased height and width (images with higher resolution) as well as images with only one dimension changed(extend and stretch).}
\label{fig:celeb_diff_sizes}
\end{figure}
\begin{figure*}[htb]
\begin{center}
\includegraphics[width=1.\textwidth]{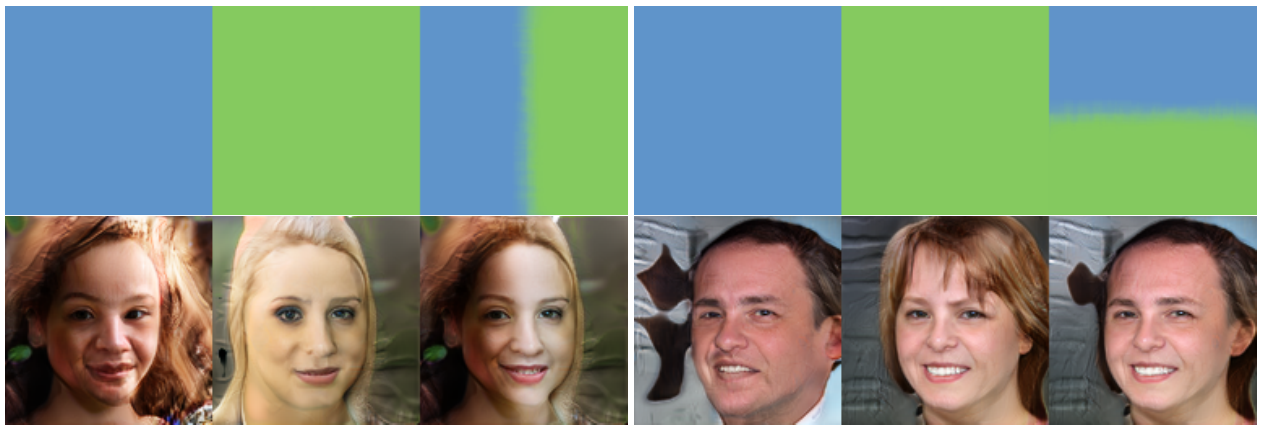} 
\end{center}
\caption{Horizontal and vertical interpolation between two images. {\bf Top}: we present which parts of master plane was taken to produce target image. 
{\bf Bottom:} we visualize original images and a produce of mixing master plan features. 
}
\label{fig:celeb_2_inter}
\end{figure*}

Observe that since our model uses local learning, the discriminator never sees the whole patch, and consequently has no information that the data was not periodic. 
Since we have periodic position channels, by using periodic local channels (with the same period), our model will produce periodic patterns.
In Fig.~\ref{fig:peridic_2} we present original images (in first row) element generated by our model (we highlight the part which is periodic). 

Because the network is fully convolutional, a semi-periodic pattern could be generated based on the network trained on one image alone. Simplest way to achieve that is to use different local plane for each periodic element. In the last three rows in Fig.~\ref{fig:peridic_2} we show that we can change local plan to provide a little modification for each periodic element. In consequence we obtain a semi-periodic pattern.

We can also use double periodic coordinates on X and Y position to produce images with arbitrary size, see Fig.~\ref{fig:peridic_int}.

\subsection{FFHQ}

In the case of FFHQ data set our model is trained on  ($64 \times 64$) randomly cropped elements of original data set ($128 \times 128$). In Fig.~\ref{fig:celeba_samples}  we present images ($128 \times 128$)  generated by \our{} model.

In our model input noise layer has local and master plan channels. Master plane codes most information about the photo. Interpolation between both plans and interpolation with fixed local plan are almost the same. It may seem that a local plan is not needed, but our experiments show that it allows us to obtain sharper images and model small details such as hair structure, eye color or background. In each row of Fig.~\ref{fig:celeba-random_local} we present single face with randomly changed local plan. As we can see, local plan change affects the structure of hair.

\paragraph{Information coded by position channels}

Now we would like to perform the experiment which shows that the use of position channels is crucial. 
In the left upper corner in Fig.~\ref{fig:celeb_diff_sizes} we have an image generate in the same resolution as the element in the training data set. We train model on resolution ($128\times128$) by using small cropped part of size ($64\times64$). During training we use coordinate of the part of an image because model should know which part of the image we used. After training we can generate elements of size ($128\times128$).

In our model it is also possible to produce images with larger resolution than elements from training data set. We obtain it by extending an array of coordinates with bilinear interpolation. If we increase length and width, we obtain original image with higher resolution. On the other hand we can modify only one edge of the image and extend or stretch the face, see Fig.~\ref{fig:celeb_diff_sizes}.


\paragraph{Transfer}

Master plan channels encode global features of images. 
Since our construction is fully convolutional, we have no guarantee that, for example, the left upper local corner of the input would be able to exchange information with the right lower. In consequence, for the FFHQ data set we could theoretically obtain a photo where the left side of the image comes from a man, while the right side comes from a woman. Thus, we need to insert global information. We do this by master-plan channels.

Observe that although obviously some features have strong correlation with location, in the typical GAN models this correspondence is lost. And to modify some features locally one would typically need some form of disentanglement.

Contrary to this, in our model information which corresponds to the position is embedded in the representation. This allows to ,,transplant'' hair, or glasses/eye(s), nose or beard from one face to another. In particular, we can easily generate a face with different eye colors -- which clearly goes out of the distribution of the data.

To some extent this resembles the style transfer. In the first experiments we take two generated images and produce one which on one half contains master plan from one image and in the second half has a master plan from another one, see Fig.~\ref{fig:celeb_2_inter}. We construct two approaches: we merge images horizontally and vertically. As we can see, the  new image contains a smooth interpolation between two source faces.

We can also transfer style only with some particular part of an image. In Fig.~\ref{fig:celeba_attr} we present how to transfer smiles and glasses.

\section{Conclusion}

The crucial aspect of \our{} is the use of a latent consisting of noise-like images with possibly different resolutions and the input made of position, local and global channels. Moreover, the noise-like images of different resolutions are processed locally.

Although the architecture of \our{} is simple, it has some important advantages over the standard GAN models. In particular, we can produce images of various sizes, learn on just one image with a pattern, construct periodic images, or ,,transplant'' parts of images.

\section{Acknowledgements}

The work of P. Spurek was supported by the National Centre of Science (Poland) Grant No. 2019/33/B/ST6/00894. The work of \L. Struski was supported by the National Centre of Science (Poland) Grant No. 2020/39/D/ST6/01332.
The research of S. Knop was funded by the Priority Research Area Digiworld under the program Excellence Initiative – Research University at the Jagiellonian University in Kraków
J. Tabor carried out this work within the research project ``Bio-inspired artificial neural network'' (grant no. POIR.04.04.00-00-14DE/18-00) within the Team-Net program of the Foundation for Polish Science co-financed by the European Union under the European Regional Development Fund.

\bibliographystyle{icml2020}

\begin{thebibliography}{34}
\providecommand{\natexlab}[1]{#1}
\providecommand{\url}[1]{\texttt{#1}}
\expandafter\ifx\csname urlstyle\endcsname\relax
  \providecommand{\doi}[1]{doi: #1}\else
  \providecommand{\doi}{doi: \begingroup \urlstyle{rm}\Url}\fi

\bibitem[Arjovsky et~al.(2017)Arjovsky, Chintala, and
  Bottou]{arjovsky2017wasserstein}
Arjovsky, M., Chintala, S., and Bottou, L.
\newblock Wasserstein generative adversarial networks.
\newblock In \emph{International conference on machine learning}, pp.\
  214--223, 2017.

\bibitem[Bahdanau et~al.(2014)Bahdanau, Cho, and Bengio]{bahdanau2014neural}
Bahdanau, D., Cho, K., and Bengio, Y.
\newblock Neural machine translation by jointly learning to align and
  translate.
\newblock \emph{arXiv preprint arXiv:1409.0473}, 2014.

\bibitem[Bellemare et~al.(2017)Bellemare, Danihelka, Dabney, Mohamed,
  Lakshminarayanan, Hoyer, and Munos]{bellemare2017cramer}
Bellemare, M.~G., Danihelka, I., Dabney, W., Mohamed, S., Lakshminarayanan, B.,
  Hoyer, S., and Munos, R.
\newblock The cramer distance as a solution to biased wasserstein gradients.
\newblock \emph{arXiv preprint arXiv:1705.10743}, 2017.

\bibitem[Cheng et~al.(2016)Cheng, Dong, and Lapata]{cheng2016long}
Cheng, J., Dong, L., and Lapata, M.
\newblock Long short-term memory-networks for machine reading.
\newblock \emph{arXiv preprint arXiv:1601.06733}, 2016.

\bibitem[Dinh et~al.(2014)Dinh, Krueger, and Bengio]{dinh2014nice}
Dinh, L., Krueger, D., and Bengio, Y.
\newblock {NICE}: Non-linear independent components estimation.
\newblock \emph{arXiv:1410.8516}, 2014.

\bibitem[Goodfellow et~al.(2014)Goodfellow, Pouget-Abadie, Mirza, Xu,
  Warde-Farley, Ozair, Courville, and Bengio]{goodfellow2014generative}
Goodfellow, I., Pouget-Abadie, J., Mirza, M., Xu, B., Warde-Farley, D., Ozair,
  S., Courville, A., and Bengio, Y.
\newblock Generative adversarial nets.
\newblock In \emph{Advances in neural information processing systems}, pp.\
  2672--2680, 2014.

\bibitem[Gulrajani et~al.(2017)Gulrajani, Ahmed, Arjovsky, Dumoulin, and
  Courville]{gulrajani2017improved}
Gulrajani, I., Ahmed, F., Arjovsky, M., Dumoulin, V., and Courville, A.~C.
\newblock Improved training of wasserstein gans.
\newblock In \emph{Advances in neural information processing systems}, pp.\
  5767--5777, 2017.

\bibitem[Heusel et~al.(2017)Heusel, Ramsauer, Unterthiner, Nessler, and
  Hochreiter]{heusel2017gans}
Heusel, M., Ramsauer, H., Unterthiner, T., Nessler, B., and Hochreiter, S.
\newblock Gans trained by a two time-scale update rule converge to a local nash
  equilibrium.
\newblock In \emph{Advances in Neural Information Processing Systems}, pp.\
  6626--6637, 2017.

\bibitem[Hong et~al.(2018)Hong, Yang, Choi, and Lee]{hong2018inferring}
Hong, S., Yang, D., Choi, J., and Lee, H.
\newblock Inferring semantic layout for hierarchical text-to-image synthesis.
\newblock In \emph{Proceedings of the IEEE Conference on Computer Vision and
  Pattern Recognition}, pp.\  7986--7994, 2018.

\bibitem[Isola et~al.(2017)Isola, Zhu, Zhou, and Efros]{isola2017image}
Isola, P., Zhu, J.-Y., Zhou, T., and Efros, A.~A.
\newblock Image-to-image translation with conditional adversarial networks.
\newblock In \emph{Proceedings of the IEEE conference on computer vision and
  pattern recognition}, pp.\  1125--1134, 2017.

\bibitem[Karras et~al.(2017)Karras, Aila, Laine, and
  Lehtinen]{karras2017progressive}
Karras, T., Aila, T., Laine, S., and Lehtinen, J.
\newblock Progressive growing of gans for improved quality, stability, and
  variation.
\newblock \emph{arXiv preprint arXiv:1710.10196}, 2017.

\bibitem[Karras et~al.(2019)Karras, Laine, and Aila]{karras2019style}
Karras, T., Laine, S., and Aila, T.
\newblock A style-based generator architecture for generative adversarial
  networks.
\newblock In \emph{Proceedings of the IEEE Conference on Computer Vision and
  Pattern Recognition}, pp.\  4401--4410, 2019.

\bibitem[Kingma \& Welling(2014)Kingma and Welling]{kingma2014auto}
Kingma, D. and Welling, M.
\newblock Auto-encoding variational bayes.
\newblock \emph{arXiv:1312.6114}, 2014.

\bibitem[Kingma \& Dhariwal(2018)Kingma and Dhariwal]{kingma2018glow}
Kingma, D.~P. and Dhariwal, P.
\newblock Glow: Generative flow with invertible 1x1 convolutions.
\newblock In \emph{Advances in Neural Information Processing Systems}, pp.\
  10236--10245, 2018.

\bibitem[Kodali et~al.(2017)Kodali, Abernethy, Hays, and
  Kira]{kodali2017convergence}
Kodali, N., Abernethy, J., Hays, J., and Kira, Z.
\newblock On convergence and stability of gans.
\newblock \emph{arXiv preprint arXiv:1705.07215}, 2017.

\bibitem[Ledig et~al.(2017)Ledig, Theis, Husz{\'a}r, Caballero, Cunningham,
  Acosta, Aitken, Tejani, Totz, Wang, et~al.]{ledig2017photo}
Ledig, C., Theis, L., Husz{\'a}r, F., Caballero, J., Cunningham, A., Acosta,
  A., Aitken, A., Tejani, A., Totz, J., Wang, Z., et~al.
\newblock Photo-realistic single image super-resolution using a generative
  adversarial network.
\newblock In \emph{Proceedings of the IEEE conference on computer vision and
  pattern recognition}, pp.\  4681--4690, 2017.

\bibitem[Lin et~al.(2019)Lin, Chang, Chen, Juan, Wei, and Chen]{lin2019coco}
Lin, C.~H., Chang, C.-C., Chen, Y.-S., Juan, D.-C., Wei, W., and Chen, H.-T.
\newblock Coco-gan: generation by parts via conditional coordinating.
\newblock In \emph{Proceedings of the IEEE International Conference on Computer
  Vision}, pp.\  4512--4521, 2019.

\bibitem[Mao et~al.(2017)Mao, Li, Xie, Lau, Wang, and
  Paul~Smolley]{mao2017least}
Mao, X., Li, Q., Xie, H., Lau, R.~Y., Wang, Z., and Paul~Smolley, S.
\newblock Least squares generative adversarial networks.
\newblock In \emph{Proceedings of the IEEE International Conference on Computer
  Vision}, pp.\  2794--2802, 2017.

\bibitem[Miyato \& Koyama(2018)Miyato and Koyama]{miyato2018cgans}
Miyato, T. and Koyama, M.
\newblock cgans with projection discriminator.
\newblock \emph{arXiv preprint arXiv:1802.05637}, 2018.

\bibitem[Miyato et~al.(2018)Miyato, Kataoka, Koyama, and
  Yoshida]{miyato2018spectral}
Miyato, T., Kataoka, T., Koyama, M., and Yoshida, Y.
\newblock Spectral normalization for generative adversarial networks.
\newblock \emph{arXiv preprint arXiv:1802.05957}, 2018.

\bibitem[Parikh et~al.(2016)Parikh, T{\"a}ckstr{\"o}m, Das, and
  Uszkoreit]{parikh2016decomposable}
Parikh, A.~P., T{\"a}ckstr{\"o}m, O., Das, D., and Uszkoreit, J.
\newblock A decomposable attention model for natural language inference.
\newblock \emph{arXiv preprint arXiv:1606.01933}, 2016.

\bibitem[Park et~al.(2019)Park, Liu, Wang, and Zhu]{park2019semantic}
Park, T., Liu, M.-Y., Wang, T.-C., and Zhu, J.-Y.
\newblock Semantic image synthesis with spatially-adaptive normalization.
\newblock In \emph{Proceedings of the IEEE Conference on Computer Vision and
  Pattern Recognition}, pp.\  2337--2346, 2019.

\bibitem[Parmar et~al.(2018)Parmar, Vaswani, Uszkoreit, Kaiser, Shazeer, Ku,
  and Tran]{parmar2018image}
Parmar, N., Vaswani, A., Uszkoreit, J., Kaiser, {\L}., Shazeer, N., Ku, A., and
  Tran, D.
\newblock Image transformer.
\newblock \emph{arXiv preprint arXiv:1802.05751}, 2018.

\bibitem[Radford et~al.(2015)Radford, Metz, and
  Chintala]{radford2015unsupervised}
Radford, A., Metz, L., and Chintala, S.
\newblock Unsupervised representation learning with deep convolutional
  generative adversarial networks.
\newblock \emph{arXiv preprint arXiv:1511.06434}, 2015.

\bibitem[Reed et~al.(2016)Reed, Akata, Yan, Logeswaran, Schiele, and
  Lee]{reed2016generative}
Reed, S., Akata, Z., Yan, X., Logeswaran, L., Schiele, B., and Lee, H.
\newblock Generative adversarial text to image synthesis.
\newblock \emph{arXiv preprint arXiv:1605.05396}, 2016.

\bibitem[S{\o}nderby et~al.(2016)S{\o}nderby, Caballero, Theis, Shi, and
  Husz{\'a}r]{sonderby2016amortised}
S{\o}nderby, C.~K., Caballero, J., Theis, L., Shi, W., and Husz{\'a}r, F.
\newblock Amortised map inference for image super-resolution.
\newblock \emph{arXiv preprint arXiv:1610.04490}, 2016.

\bibitem[Taigman et~al.(2016)Taigman, Polyak, and
  Wolf]{taigman2016unsupervised}
Taigman, Y., Polyak, A., and Wolf, L.
\newblock Unsupervised cross-domain image generation.
\newblock \emph{arXiv preprint arXiv:1611.02200}, 2016.

\bibitem[Tolstikhin et~al.(2017)Tolstikhin, Bousquet, Gelly, and
  Schoelkopf]{tolstikhin2017wasserstein}
Tolstikhin, I., Bousquet, O., Gelly, S., and Schoelkopf, B.
\newblock Wasserstein auto-encoders.
\newblock \emph{arXiv:1711.01558}, 2017.

\bibitem[Vaswani et~al.(2017)Vaswani, Shazeer, Parmar, Uszkoreit, Jones, Gomez,
  Kaiser, and Polosukhin]{vaswani2017attention}
Vaswani, A., Shazeer, N., Parmar, N., Uszkoreit, J., Jones, L., Gomez, A.~N.,
  Kaiser, {\L}., and Polosukhin, I.
\newblock Attention is all you need.
\newblock In \emph{Advances in neural information processing systems}, pp.\
  5998--6008, 2017.

\bibitem[Xu et~al.(2015)Xu, Ba, Kiros, Cho, Courville, Salakhudinov, Zemel, and
  Bengio]{xu2015show}
Xu, K., Ba, J., Kiros, R., Cho, K., Courville, A., Salakhudinov, R., Zemel, R.,
  and Bengio, Y.
\newblock Show, attend and tell: Neural image caption generation with visual
  attention.
\newblock In \emph{International conference on machine learning}, pp.\
  2048--2057, 2015.

\bibitem[Yu et~al.(2015)Yu, Seff, Zhang, Song, Funkhouser, and
  Xiao]{yu2015lsun}
Yu, F., Seff, A., Zhang, Y., Song, S., Funkhouser, T., and Xiao, J.
\newblock Lsun: Construction of a large-scale image dataset using deep learning
  with humans in the loop.
\newblock \emph{arXiv preprint arXiv:1506.03365}, 2015.

\bibitem[Zhang et~al.(2018{\natexlab{a}})Zhang, Goodfellow, Metaxas, and
  Odena]{zhang2018self}
Zhang, H., Goodfellow, I., Metaxas, D., and Odena, A.
\newblock Self-attention generative adversarial networks.
\newblock \emph{arXiv preprint arXiv:1805.08318}, 2018{\natexlab{a}}.

\bibitem[Zhang et~al.(2018{\natexlab{b}})Zhang, Xu, Li, Zhang, Wang, Huang, and
  Metaxas]{zhang2018stackgan++}
Zhang, H., Xu, T., Li, H., Zhang, S., Wang, X., Huang, X., and Metaxas, D.~N.
\newblock Stackgan++: Realistic image synthesis with stacked generative
  adversarial networks.
\newblock \emph{IEEE transactions on pattern analysis and machine
  intelligence}, 41\penalty0 (8):\penalty0 1947--1962, 2018{\natexlab{b}}.

\bibitem[Zhu et~al.(2017)Zhu, Park, Isola, and Efros]{zhu2017unpaired}
Zhu, J.-Y., Park, T., Isola, P., and Efros, A.~A.
\newblock Unpaired image-to-image translation using cycle-consistent
  adversarial networks.
\newblock In \emph{Proceedings of the IEEE international conference on computer
  vision}, pp.\  2223--2232, 2017.

\end{thebibliography}

\section{Appendix A: Generating samples with different sizes}
Despite the fact, entire images are never used in training,
\our{} produces full, state-of-the-art-quality images.
We further demonstrate a variety of novel applications enabled by teaching
the network to be aware of coordinates as well as local/global latent. In Fig.~\ref{fig:lsun_resize} we present samples with different
sizes, generated by a model trained on LSUN (bedrooms)
data set.

\section{Appendix B: Information coded by position channels}
\begin{figure*}[htb]
	\centering
	\includegraphics[width=1.\textwidth]{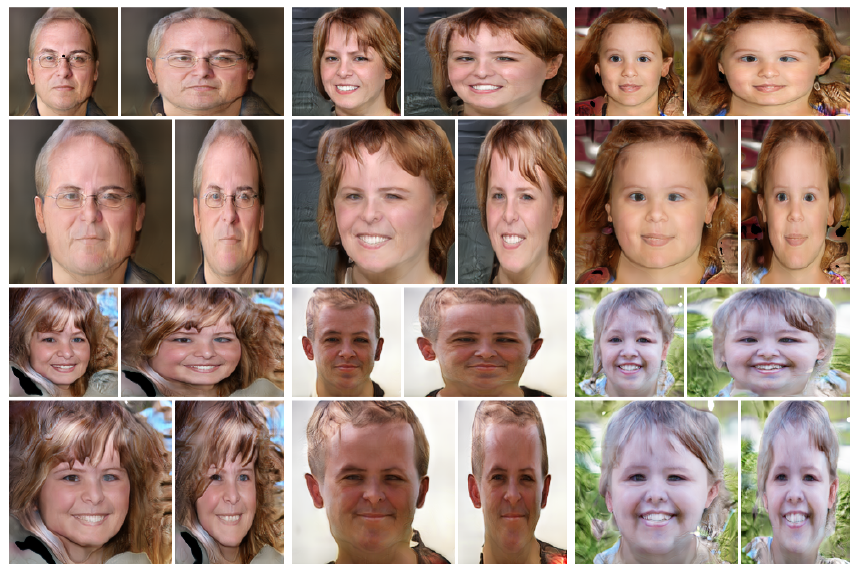}
	\caption{The picture presents samples with different sizes that were generated from a model trained on the FFHQ data set. Each sample consists of images with resolutions $128\times128$, $128\times192$ in the upper row and  $192\times128$, $192\times192$ in the row below.}
	\label{fig:celeba_diff_size}
\end{figure*}

In this paragraph, we would like to show more experiments that present the potential use of position channels. In the left upper corner in
Fig.~\ref{fig:celeba_diff_size} we have an image generated in the same resolution
as the element in the training data set. We train the model on
resolution ($128\times128$) by using small cropped part of size
($64\times64$). During training we use coordinate of the part
of an image because model should know which part of the
image we used. After training, we can generate elements of
size ($128\times128$).

%
\section{Appendix C: Transfer of style}

\begin{figure*}[htb]
	\centering
	\includegraphics[width=1.\textwidth]{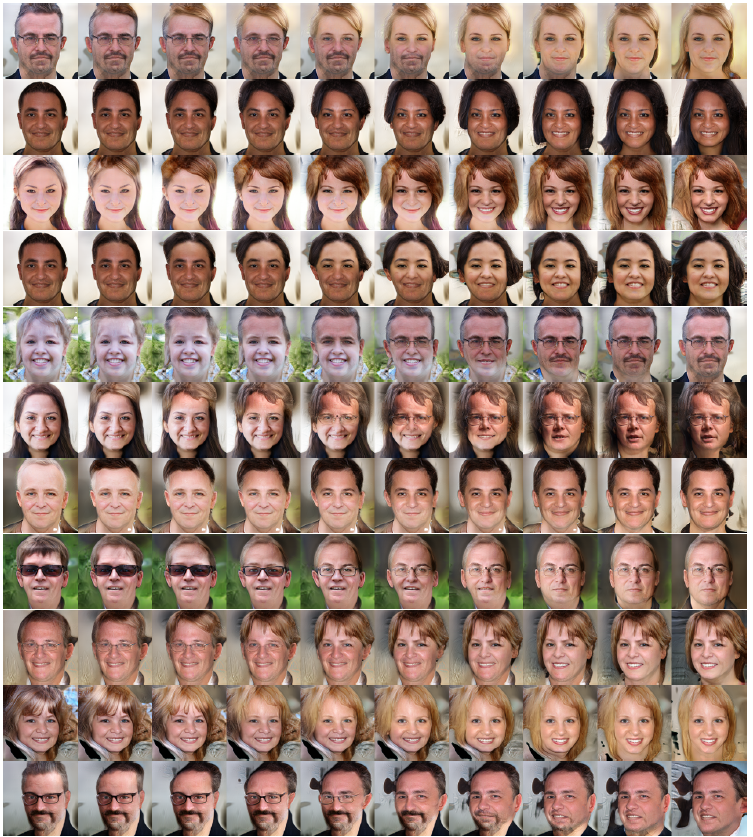}
	\caption{Vertical (from top to bottom) interpolation between two images.}
	\label{fig:celeba_curtain}
\end{figure*}

In \our{} information that corresponds
to the position is embedded in the representation.
In the main paper, we show experiment when we take two generated images and produce one
which one half contains master plan from one image and
in the second half has a master plan from another one, see
Fig. 11. Below we present a modification of that experiment where we construct linear interpolation between two images by vertically swapping master plan from one image to another. In the beginning, we have one image which is changing (from top to bottom) into another, see Fig.~\ref{fig:celeba_curtain}.

We can also transfer style only with some particular part of
an image. In Fig.~\ref{fig:celeba_attr} we present how to transfer smiles and
glasses. In the first row we have images, which are changed by adding some elements from images in the first column.

%
%
%
%
%
%

\begin{figure*}[htb]
	\centering
	\includegraphics[width=1.\textwidth]{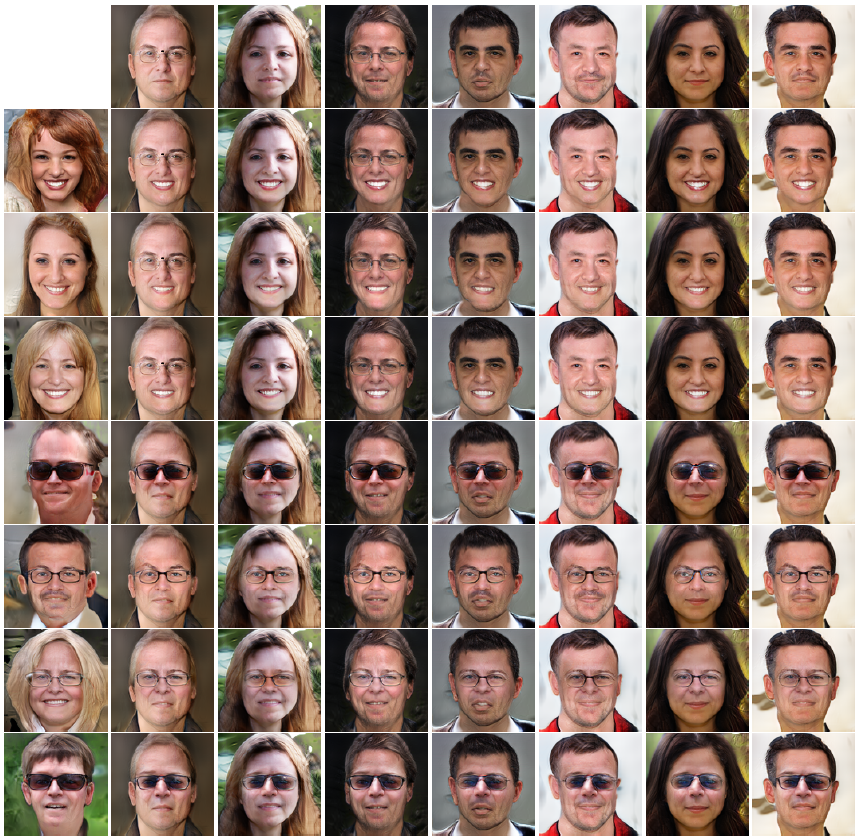}
	\caption{Moving attributes from one image to another. In the first row, there are images, which are changed in below rows by transferring some elements from images in the first column.
    Due to the fact that our model is local, we do not change the image except for a neighborhood of the ,,transplanted” positions.}
	\label{fig:celeba_attr}
\end{figure*}


\begin{figure*}[htb]
	\centering
	\includegraphics[width=0.32\textwidth]{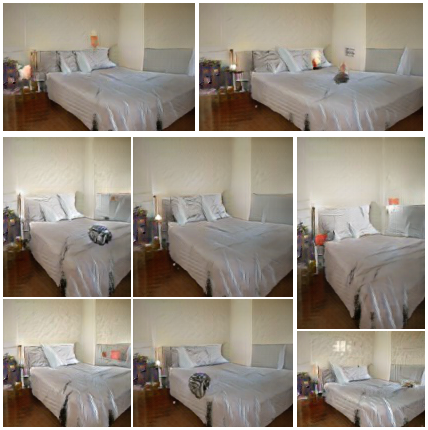}
	\includegraphics[width=0.32\textwidth]{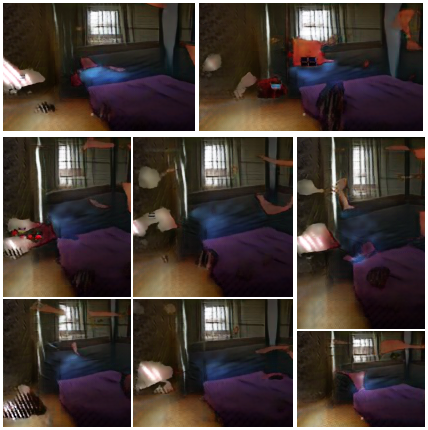}
	\includegraphics[width=0.32\textwidth]{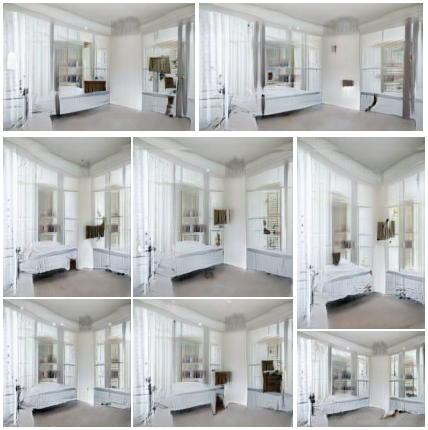}
	\caption{The picture presents samples with different sizes that were generated from a model trained on LSUN (bedrooms) data set. Each sample consists of eight images with resolutions  $128\times192$, $128\times224$ in the top row, $160\times128$, $160\times160$, $192\times128$ in the middle row and $128\times128$, $128\times160$, $96\times128$ in the lowest row}
	\label{fig:lsun_resize}
\end{figure*}

\end{document}